\documentclass[twocolumn,times, tighten]{aastex7}

\usepackage{multirow}

\newcommand{\hb}{H$\beta$}

\newcommand{\cak}{\ion{Ca}{2}~K}

\newcommand{\RN}[1]{\textup{\lowercase\expandafter{\romannumeral#1}}}

\usepackage{soul}

\shorttitle{Striations with DKIST}
\shortauthors{Kuridze et al.}

\begin{document}

\title{ \Large{The striated solar photosphere observed at 0\farcs03\, resolution}}

\correspondingauthor{D. Kuridze}
\author[0000-0003-2760-2311]{David Kuridze} 
\affiliation{National Solar Observatory, 3665 Discovery Drive, Boulder, CO 80303, USA}
\email{dkuridze@nso.edu}

\author[0000-0001-6907-9739]{Friedrich W\"oger} 
\affiliation{National Solar Observatory, 3665 Discovery Drive, Boulder, CO 80303, USA}
\email{fwoeger@nso.edu}

\author[0000-0002-2554-1351]{Han Uitenbroek} 
\affiliation{National Solar Observatory, 3665 Discovery Drive, Boulder, CO 80303, USA}
\email{huitenbroek@nso.edu}

\author[0000-0001-5850-3119]{Matthias Rempel}
\affiliation{High Altitude Observatory, NSF National Center for Atmospheric Research, Boulder, CO 80307, USA}
\email{rempel@ucar.edu}

\author[0000-0003-3147-8026]{Alexandra Tritschler} 
\affiliation{National Solar Observatory, 3665 Discovery Drive, Boulder, CO 80303, USA}
\email{atritschler@nso.edu}

\author[0000-0002-7213-9787]{Thomas Rimmele}
\affiliation{National Solar Observatory, 3665 Discovery Drive, Boulder, CO 80303, USA}
\email{trimmele@nso.edu}

\author[0000-0001-9352-3027 ]{Catherine Fischer} 
\affiliation{National Solar Observatory, 3665 Discovery Drive, Boulder, CO 80303, USA}{}
\affiliation{European Space Agency (ESA), European Space Astronomy Centre (ESAC),
Camino Bajo del Castillo s/n, 28692 Villanueva de la Cañada, Madrid, Spain}{}
\email{Catherine.Fischer@esa.int}

\author[0000-0003-4708-1074]{Oskar Steiner} 
\affiliation{Kiepenheuer-Institut f\"ur Sonnenphysik (KIS), Georges-K\"ohler-Allee 401a, 79110 Freiburg, Germany}{}
\affiliation{Istituto ricerche solari Aldo e Cele Dacc\`o (IRSOL), Via Patocchi 57, 6605 Locarno-Monti, Switzerland}{}
\email{steiner@leibniz-kis.de}


\begin{abstract}
Striated granular edges observed in the solar photosphere represent one of the smallest-scale phenomena on the Sun.
They arise from the interaction of strongly coupled hydrodynamic, magnetic, and radiative properties of the plasma.
In particular, modulations in the photospheric magnetic field strength cause variations in density and opacity along 
the line of sight, leading to their formation. Therefore, the striation patterns can be used as valuable diagnostics 
for studying the finest-scale structure of the photospheric magnetic field. The Daniel K. Inouye Solar Telescope allows 
observations of the solar atmosphere with a spatial resolution of better than 0\farcs03\ with its current instrumentation. 
We analyze images acquired with the Visible Broadband Imager using the G-band channel to investigate the characteristics 
of fine-scale striations in the photosphere and compare them with state-of-the-art radiation-MHD simulations 
at similar spatial resolution. Both observed and synthetic images reveal photospheric striae with widths of 
approximately 20$-$50 km, suggesting that at least 4-meter class solar telescopes are necessary to resolve this ultrafine structure. 
Analysis of the numerical simulations confirms that the striation observed in the filtergrams is associated with 
spatial variations in photospheric magnetic flux concentrations, which cause shifts in the geometrical height 
where the emergent intensity forms. Some fine-scale striation in the synthetic images originate from magnetic 
field variations of approximately a hundred Gauss, resulting in a Wilson depressions as narrow as 10 km. 
This suggests that DKIST G-band images can trace footprints of magnetic field variations and Wilson depressions 
at a similar scale.
\end{abstract}



\section{Introduction}
\label{intro}

\begin{figure*}
\centering
\includegraphics[width=18.0cm]{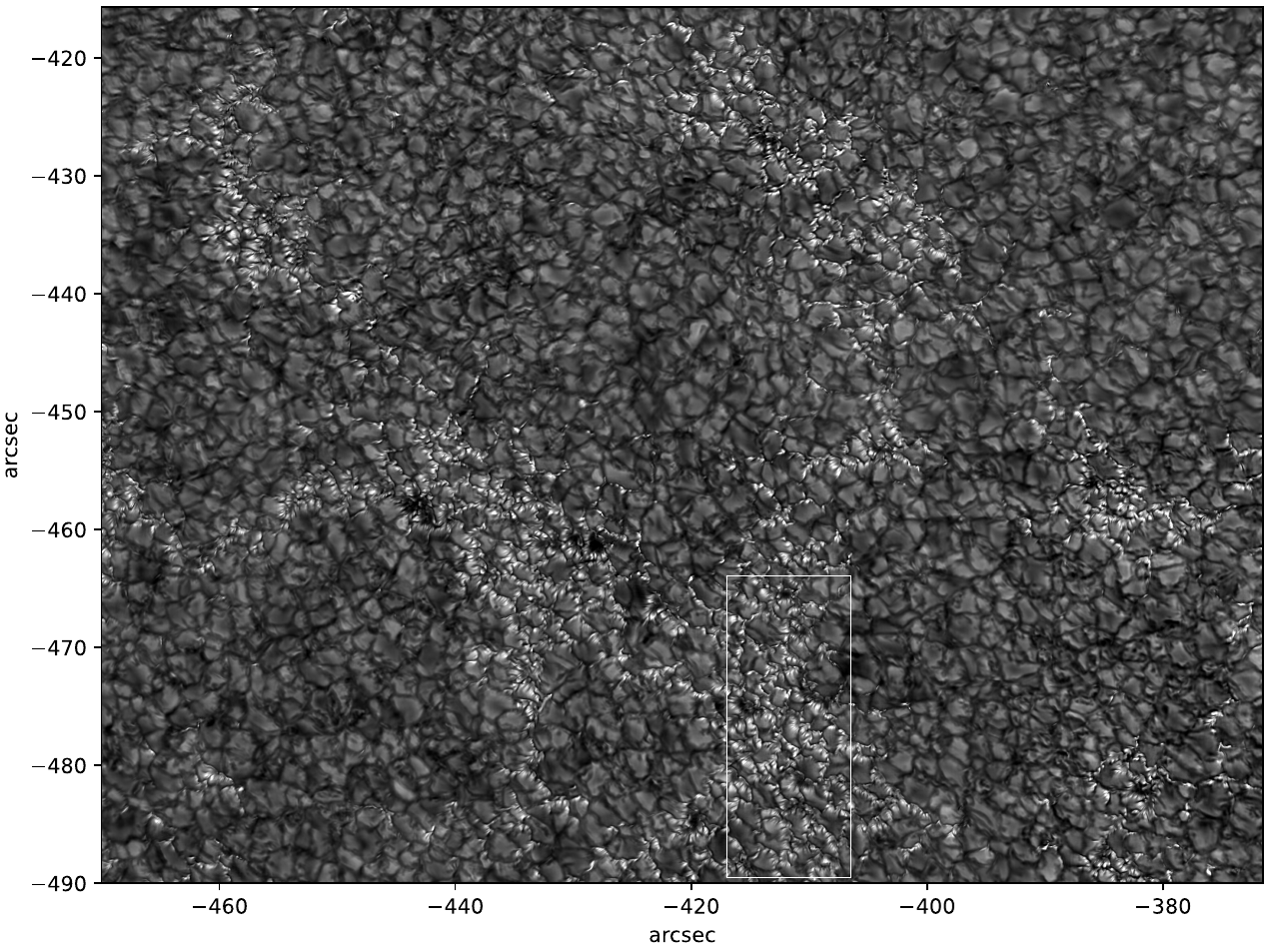}
\caption{Mosaic of high-resolution G-band filtergrams obtained with
the VBI instrument on 2022 June 3 between 17:46 UT. The white box indicates the plage 
region shown in Figure~\ref{fig2}.}
\label{fig1}
\end{figure*}

\begin{figure*}
\centering
\includegraphics[width=17.4cm]{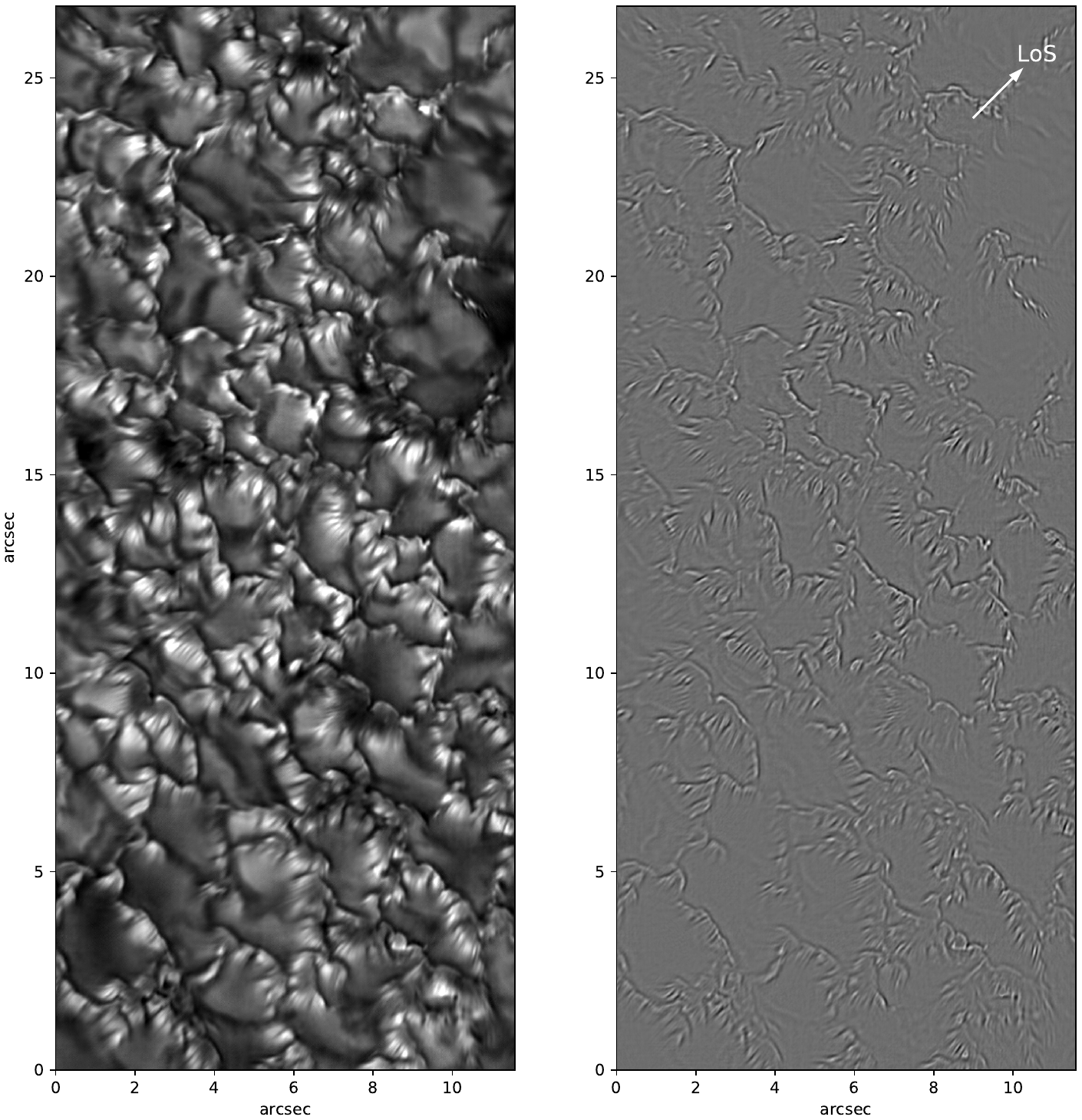}
\caption{Left: Zoomed image of the G-band facular region indicated by the white box in Figure~\ref{fig1}.
Right: The same image processed with a Gaussian filter subtraction  method (see text for an explanation).
The direction arrow points in the direction of the DC. }
\label{fig2}
\end{figure*}

Fine-scale striation in the solar photosphere is commonly observed in features known as faculae $-$ bright granular 
edges that appear in active region (AR) plages when viewed away from disk center \citep{Spruit1976,Berger1995,BergerandTitle2001,hirzberger+wiehr2005}. 
Another characteristic of faculae is the presence of a prominent dark lane on the disk-center side of the bright granular edge.

A series of excellent studies dedicated to the theoretical 
and observational aspects of faculae was published between 2004 and 
2005 \citep{Keller2004,Carlsson2004, Berger2004, Lites2004, Steiner2005}. The center-to-limb variation of G-band 
images synthesized from advanced radiative magnetoconvective simulations closely 
reproduces the main observational characteristics of the plage-type photosphere, including faculae. In particular, similar to observations, 
magnetic flux concentrations (MFCs) appear bright and with enhanced contrast in G-band images at 
or near  disk center (DC) \citep{rutten+al2001,Steiner2001,schuessler+al2003}. 
This occurs due to the low-density, low-opacity environment along the flux tube, which results in a depression 
of the optical depth unity surface where the plasma temperature is higher than outside the flux tube at the same optical depth.
Toward the limb, when MFCs are viewed at an angle, only a portion of the vertically oriented magnetic flux tubes 
intersects the line of sight (LoS), leading to the formation of facular brightenings and dark lanes  \citep{Keller2004,Carlsson2004,Steiner2005}.

The first detection of striation in photospheric faculae was made with $\sim$\,0\,\farcs1\, resolution observations 
using 1-meter class solar telescopes. First, thin dark stripes in faculae were seen in the data obtained from the Swedish Solar Telescope (SST) 
using blue continuum filtergrams centered at 487.7~nm \citep{Lites2004}. 
Later, \citet{Carlsson2004,Depontieu2006}, and \citet{Berger2007} also detected adjacent, thin, dark striations extending 
along faculae in the limbward direction, using G-band images from the SST. 

\cite{Depontieu2006} studied the temporal evolution of facular elements and concluded that the striation is a
highly dynamic phenomenon characterized by merging, splitting, and rapid motions on timescales of the order of minutes.

It has been proposed that the striation is generated by spatial variations in the magnetic field strength along the 
LoS in front of faculae \citep{Carlsson2004, Depontieu2006}. Individual dark stripes are associated with weak magnetic flux concentrations, 
while the bright space in between is due to strong magnetic fields which act as radiative channels due to their material deficiency and increased transparency.
\cite{Berger2007} showed that the striation in G-band images is closely correlated with the variation in the magnetic 
flux density measured in simultaneous photospheric magnetograms. Therefore, striation provides a unique 
opportunity to observe and investigate the small-scale structure and spatio-temporal variation of magnetic flux concentrations.
Moreover, its dynamics  offer the possibility to directly peer at the source of mechanical energy 
at the basis of magnetic flux concentrations in the photosphere. 

The width of individual striae has been estimated to be at or below the resolution limits of 1-meter-class solar telescopes.
Based on SST data, the average width of the individual strands is around 0\,\farcs1 to 0\,\farcs2 which is 
indeed close to the diffraction limit of SST \citep{Lites2004, Depontieu2006,Berger2007}.

\begin{figure*}
\centering
\includegraphics[width=16.8cm]{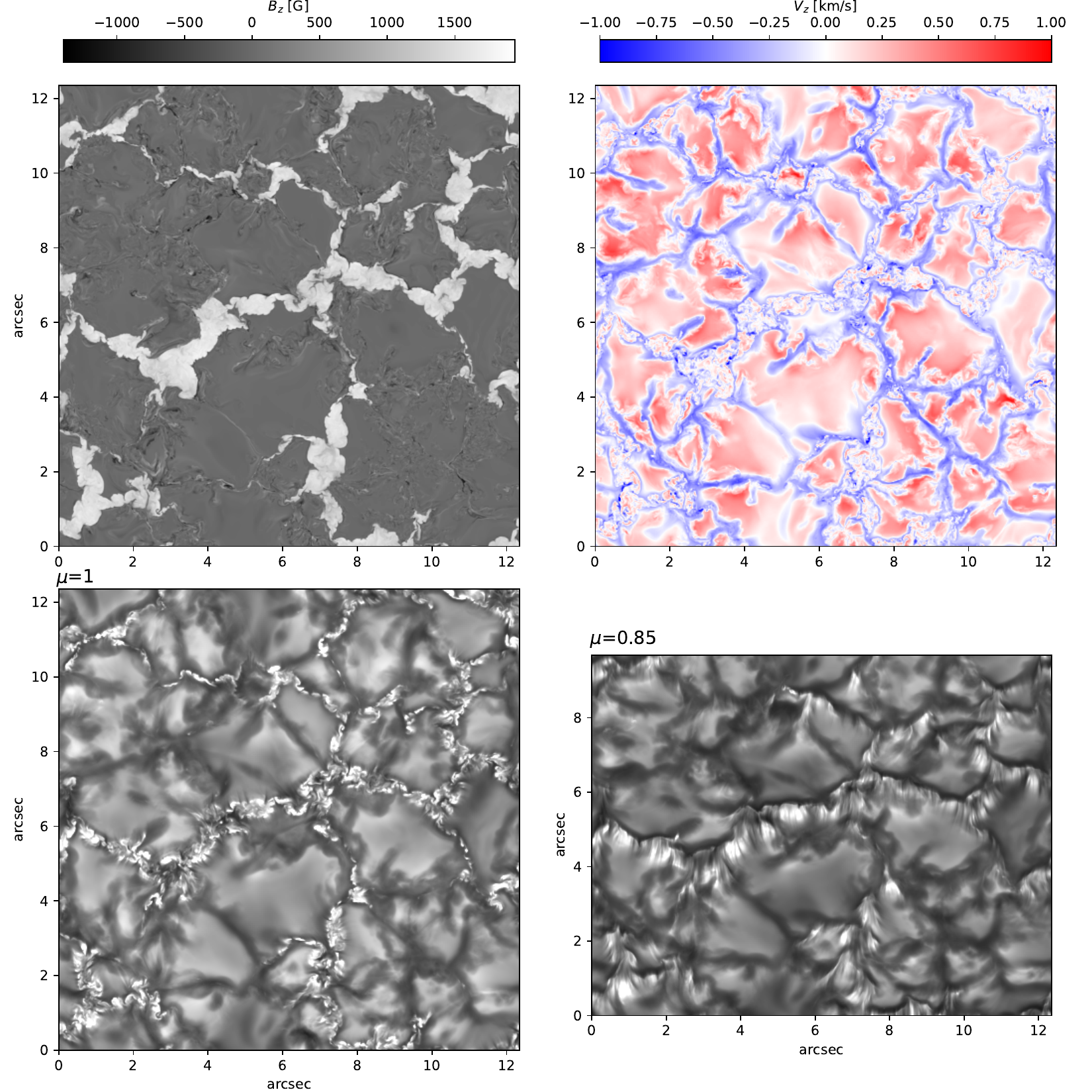}
\includegraphics[width=16.1cm]{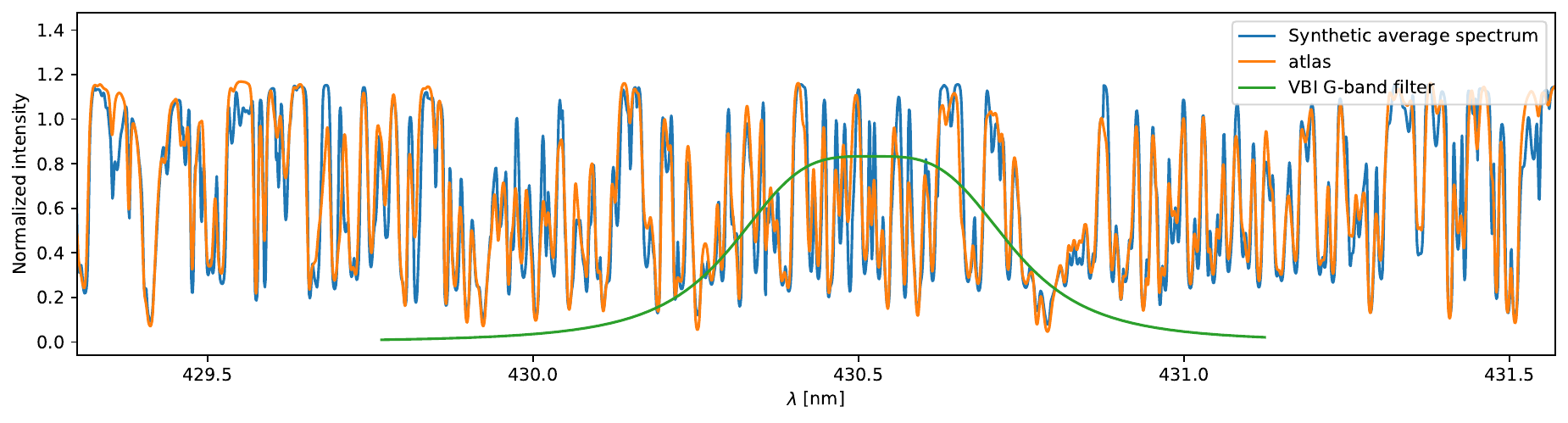}
\caption{Overview of the MURaM simulation of a plage-like atmosphere. Top row: vertical magnetic flux 
density and vertical velocity at the horizontally averaged height where the optical depth $\tau_{500\,\rm nm}$ is unity. 
Middle row: Synthetic G-band images at disk center ($\mu = 1$, left panel) and off disk center 
($\mu=0.85$, right panel) computed from the MURaM snapshot using the RH code. The limb direction is downward and the
$y$-axis is foreshortened for the limbward image in the right panel. Bottom panel:  Synthetic G-band spectrum (blue line) 
at disk center averaged over the full FoV. The orange line represents the FTS atlas profile from \cite{Neckel1984} for comparison, 
while the green line depicts the experimental transmission profile of the VBI G-band filter.}
\label{fig3}
\end{figure*}

In this work, we investigate the photospheric striation for the first time with a spatial resolution of
0\,\farcs03, using DKIST/Visible Broadband Imager data \citep{Woger2021}.
We reproduce our observations using an advanced 3D magnetohydrodynamical simulation at a 
comparable resolution, combined with spectral line synthesis, to examine the formation of
the striation on faculae and its connection with photospheric magnetic field variations.


\section{Observations}
\label{obs}

The observations used in this paper were obtained between 17:09 and 19:20 UT
on 2022 June 3 with the Visible Broadband Imager \citep[VBI;][]{Woger2021}
deployed at the Daniel K. Inouye Solar Telescope \citep[DKIST;][]{Rimmele2020}. 
The heliocentric coordinates of the center of the field-of-view (FoV) covered by the 
observations were ({\it x, y}) = ($-420$\arcsec, $-450$\arcsec), 
with heliocentric angle $\mu\approx0.85$ (see Figures~\ref{fig1}).

We used the VBI blue branch to obtain filtergrams, sequentially in G-band, \cak, and \hb\
using interference filters mounted on a rotating filterwheel. The G-band filter is centered at 430.52~nm with the 
full width half maximum (FWHM) of approximately 0.437~nm.

The instantaneous FoV of the VBI-blue is 45\arcsec$\times$45\arcsec\ with a spatial sampling of 
0\,\farcs011\,pixel$^{-1}$. For these observations, we employed the so-called ``field-sampling 
mode'' to cover the full DKIST FoV of $2^\prime\times2^\prime$ with a mosaic of 3$\times$3 tiles. 
All VBI images were reconstructed using the speckle code of \cite{Woger2008} to remove residual atmospheric 
distortion from the data, bringing the effective time step to 3~s per tile. Hence, for any given filter channel, 
the full DKIST FoV was sampled in 27 s, and the overall cadence for the three filters'
was 81 s. More details on the observations and data can be found in \cite{Kuridze2024}.

Seeing conditions during the approximately 2 hours of observation varied but were on average good throughout the observations. 
Following reconstruction, we estimate that 
multiple frames achieved a spatial resolution close to the diffraction limit (0.022\arcsec -- 0.03\arcsec at 430~nm)
based on the size of the smallest resolved structures observed in the data.

\begin{figure*}
\centering
\includegraphics[width=15.0cm]{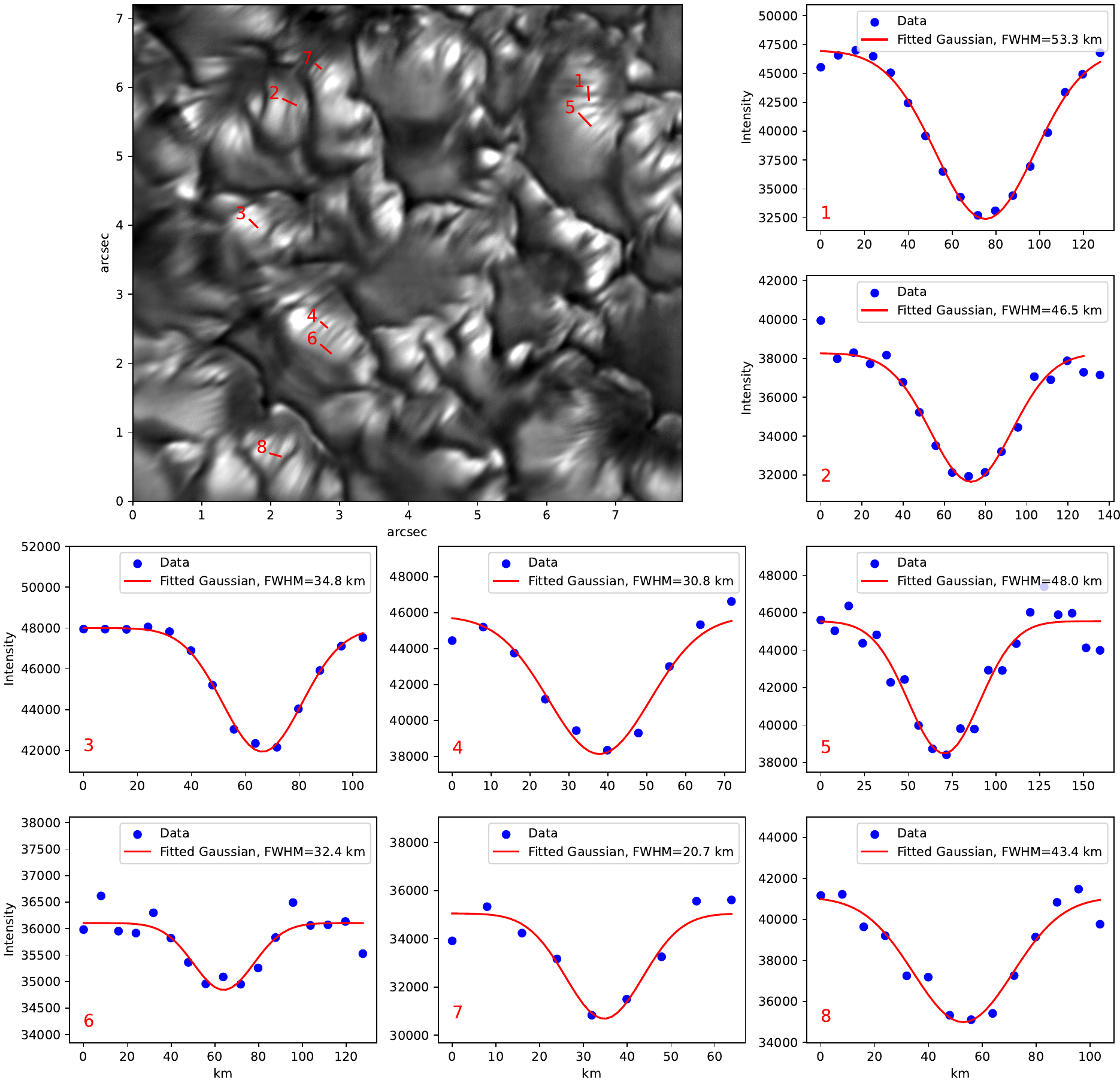}
\caption{G-band intensity profiles along the cross-cuts of dark/bright striae marked 
with red dashes in the top left observed G-band image. 
Gaussian fits of the individual profiles are overplotted in red.} 
\label{fig4}
\end{figure*}

\begin{figure*}
\centering
\includegraphics[width=17.3cm]{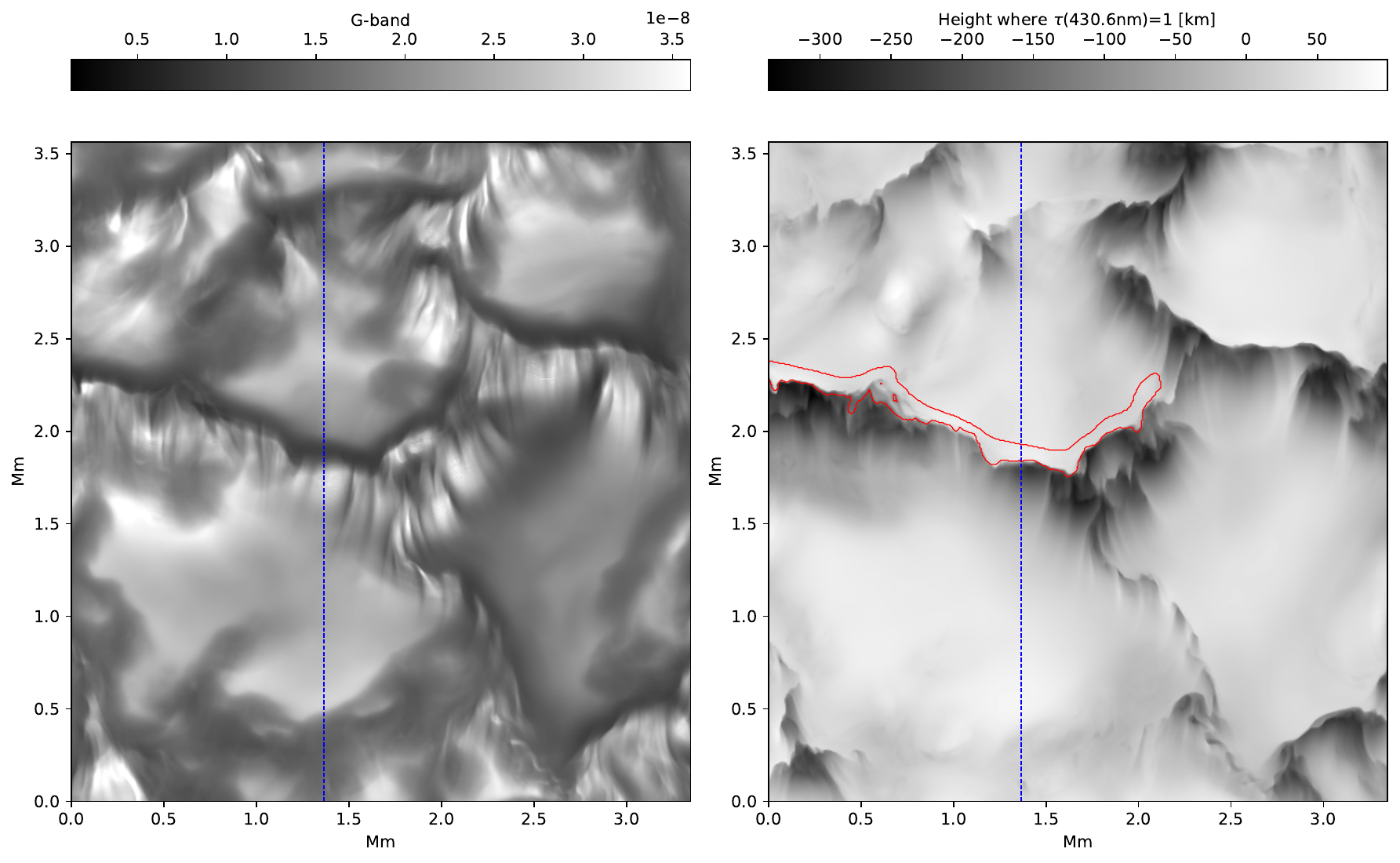}
\includegraphics[width=8.5cm]{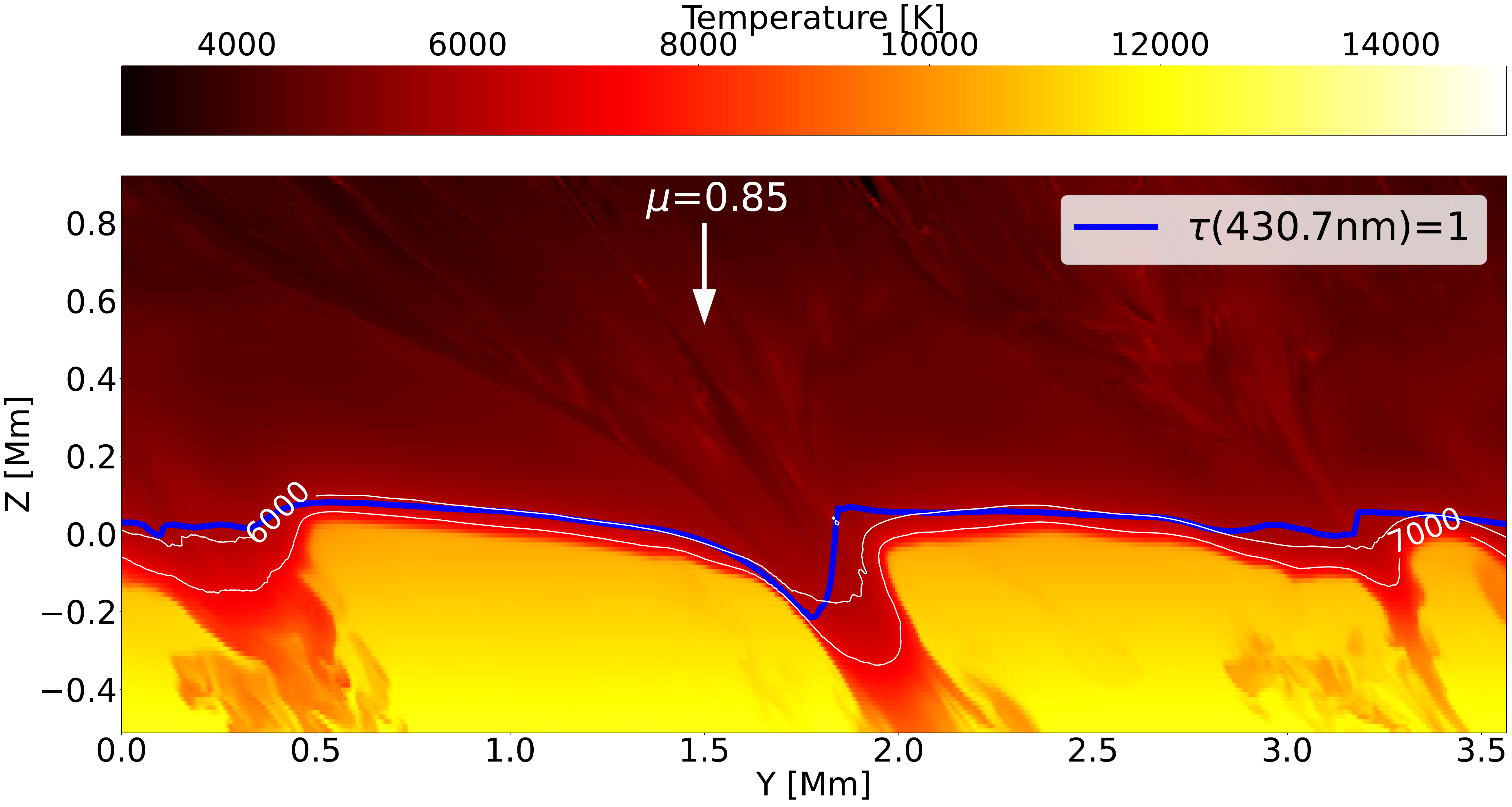}
\includegraphics[width=8.5cm]{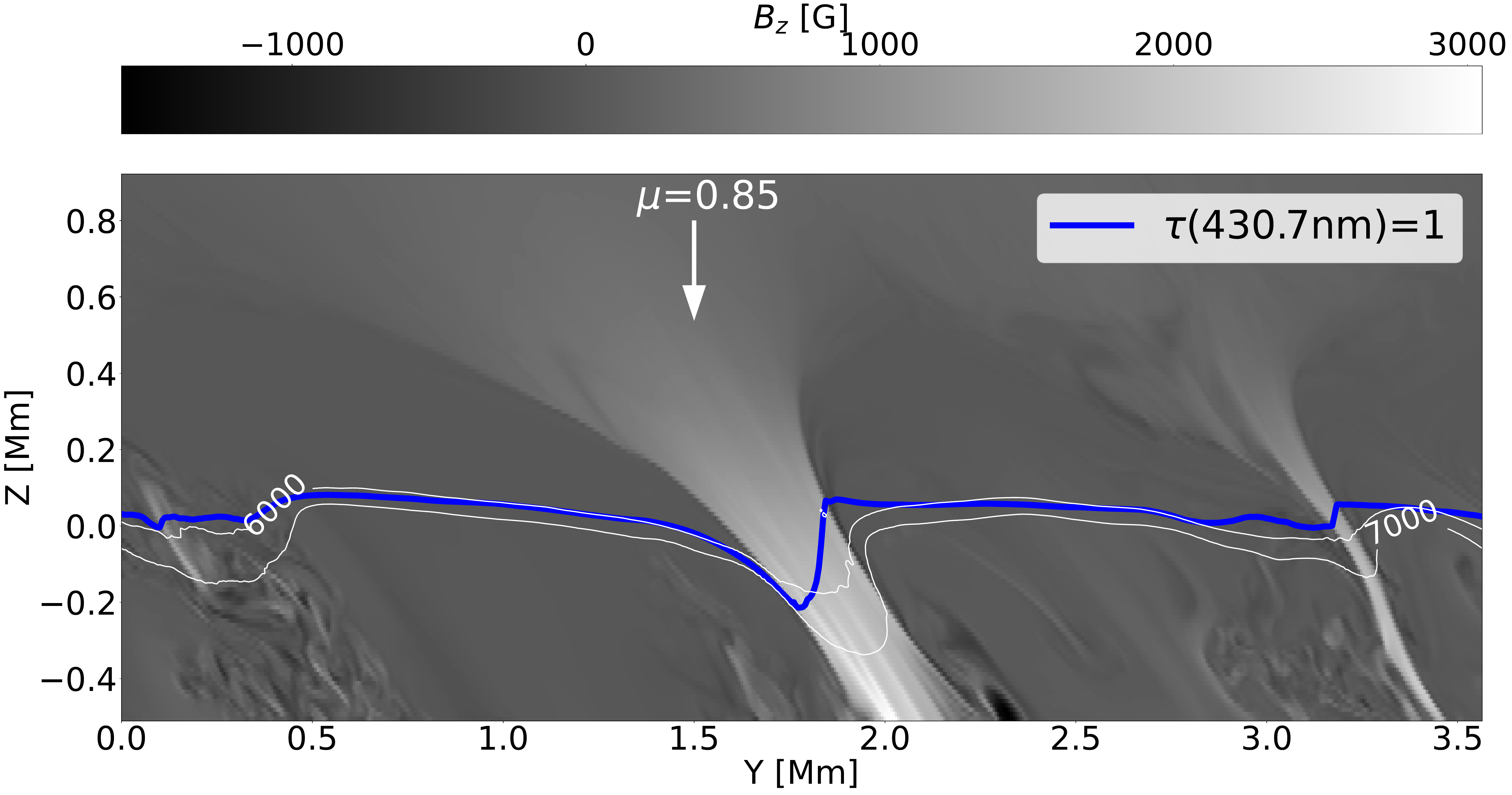}
\caption{Top: Synthetic G-band image of a simulation snapshot (left panel) and corresponding map of the height where the 
optical depth reaches unity in the G-band continuum at 430.7~nm. 
The red contour indicates the location of a dark lane starting at $x,y\approx(0, 2.4)$ in the G-band image.
The limb is in the downward direction of the images. Bottom: Vertical cuts through the temperature and LoS magnetic field 
strength along the blue dotted line marked in the top panels. 
Isothermal contours at 
$\mathrm{T=6000~and~7000~K}$ are overplotted as white lines. 
Blue lines follow the $\tau=1$ height.}
\label{fig5}
\end{figure*}

\begin{figure*}
\centering
\includegraphics[width=15.5cm]{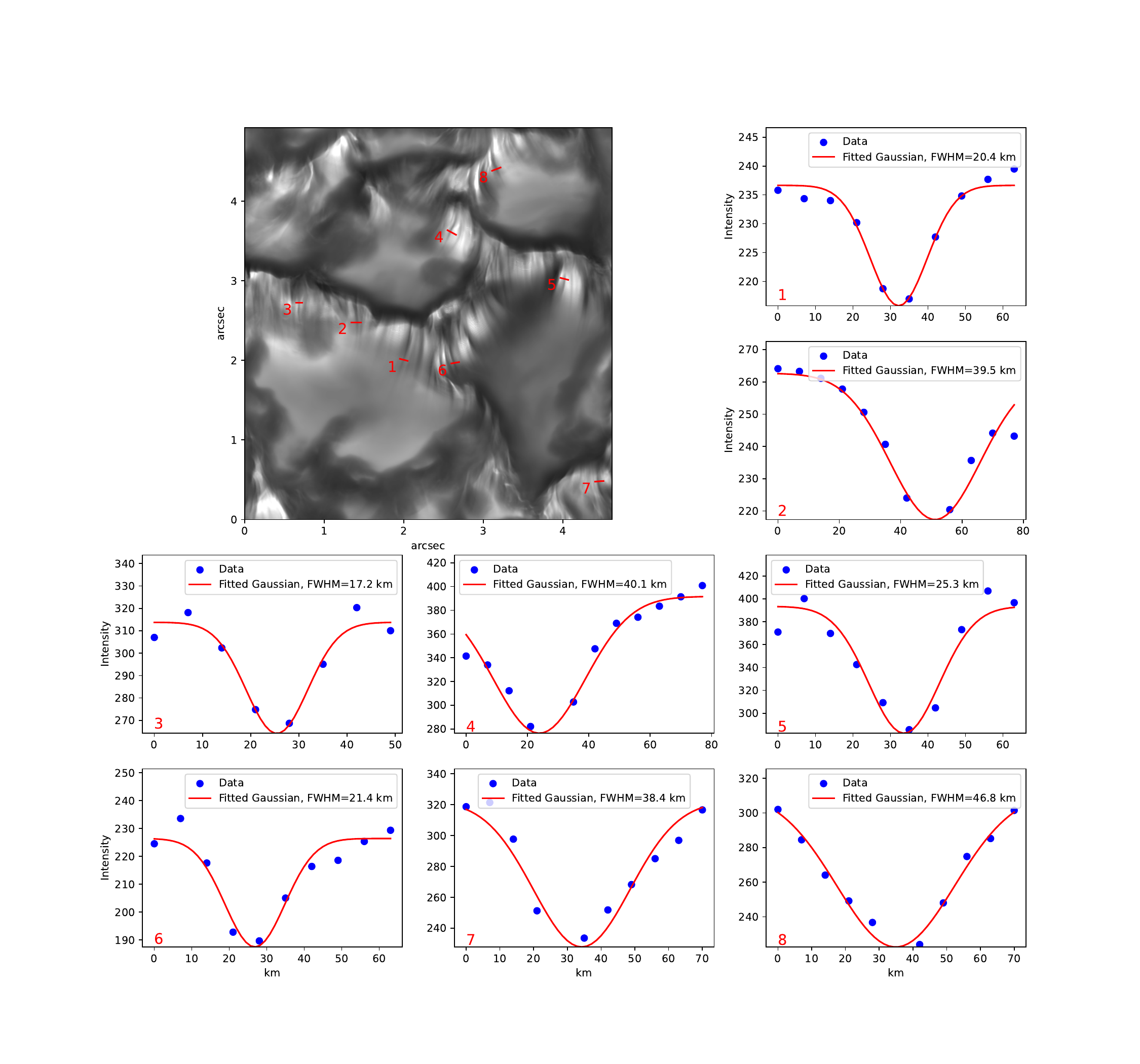}
\caption{Synthetic G-band intensity profiles along the cross-cuts of 
dark/bright striae marked with red dashes in the top left simulated G-band image computed for $\mu = 0.85$. 
Gaussian fits of the individual profiles are overplotted in red.} 
\label{fig6}
\end{figure*}


\begin{figure*}
\centering
\includegraphics[width=17.7cm]{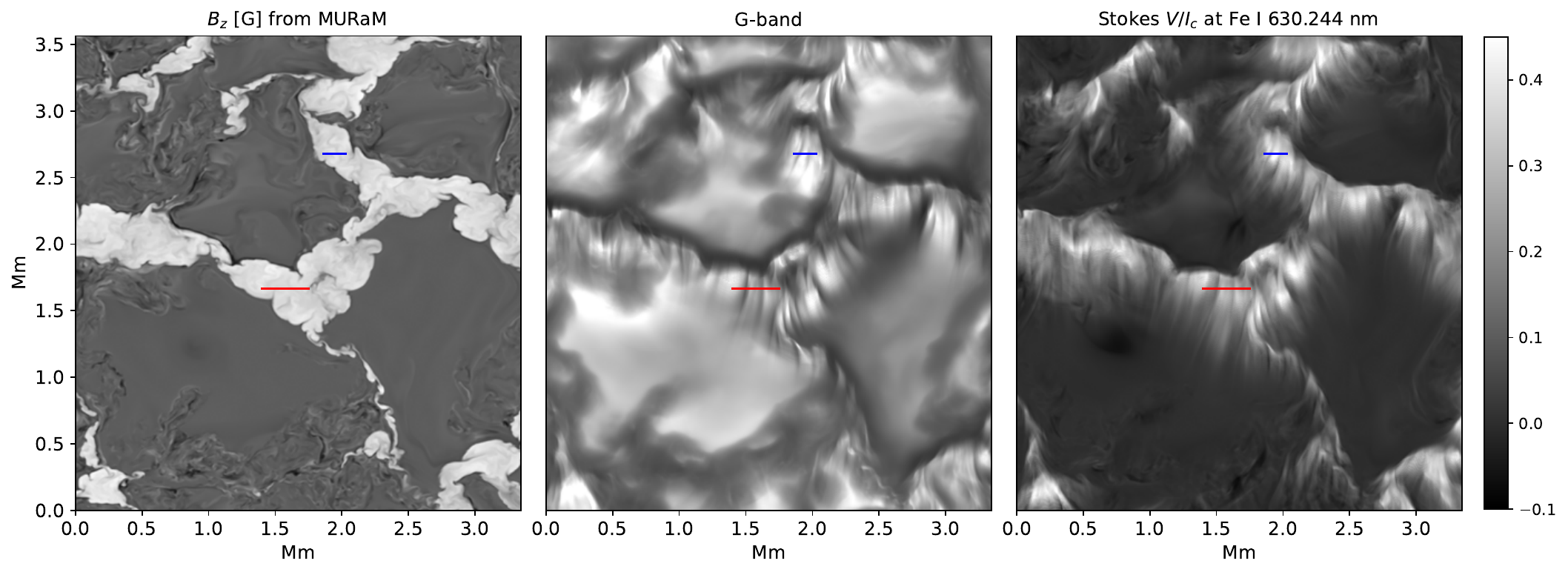}
\includegraphics[width=17.5cm]{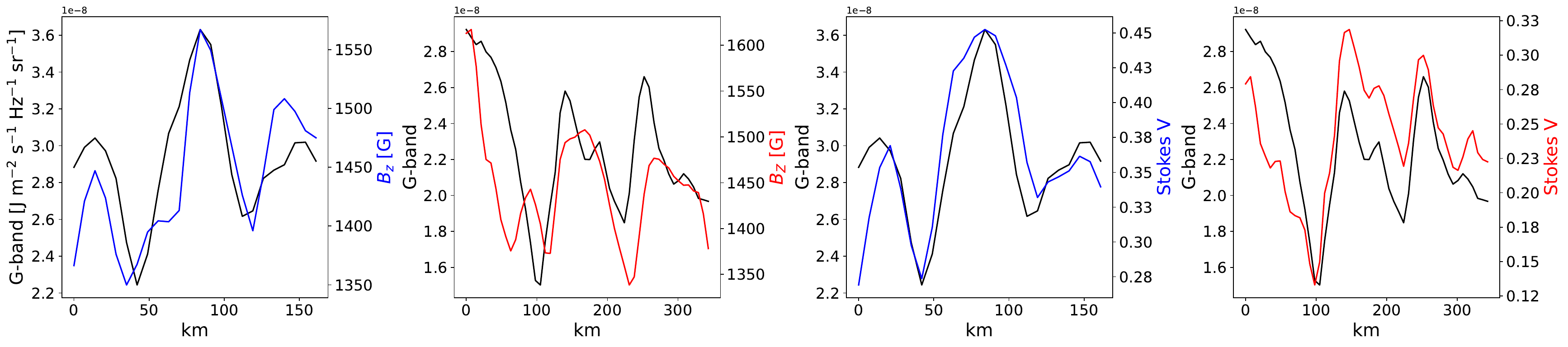}
\caption{Top: Maps of the vertical magnetic flux density, $B_z$, from the MURaM snapshot at the height $z=0$
and Stokes $V$ amplitude (blue lob) of the spectral line 
\ion{Fe}{1} 630.244~nm  (right)
normalized to the local continuum intensity. Bottom: G-band intensity, $B_z$, 
and Stokes-$V$ profiles along the red and blue lines crossing the dark striae.}
\label{fig7}
\end{figure*}



\section{Numerical Simulations and Spectral Synthesis} \label{sec:sim}

\subsection{Simulations}
\label{muram}

We carried out a numerical simulations using the radiative MHD code MURaM \citep{Rempel2014}.  
We use a setup representative of an enhanced magnetic network region with a mean net vertical magnetic flux density of 200 G,
which is based on models presented in \citet{Rempel2020} and \citet{Yeo2020}.
The absolute (unsigned) mean magnetic flux density is 260~G 
at $\tau_{500\,\rm nm}=1$ due to the presence of a small-scale dynamo.
The numerical domain has an extent of $9\times 9\times 3.24\, \mbox{Mm}^3$ (top panels of Figure~\ref{fig3}); the vertical extent covers about 2.24~Mm convection zone below the base of the photosphere ($\tau_{500\,\rm nm}=1$)
and 1~Mm of the solar atmosphere above $\tau_{500\,\rm nm}=1$. 
The side boundaries are periodic, the bottom boundary is open for convective flows and the top boundary is open for upflows, while downflows are suppressed. The magnetic bottom boundary condition allows for the transport of horizontal magnetic field \citep[see][for details]{Rempel2014} and the top boundary imposes a potential field. 

The original simulation in \citet{Rempel2020} and \citet{Yeo2020} started from a small-scale dynamo setup with an added uniform vertical magnetic field of 200 G and was evolved for 18 hours. We continued the simulation for 4000 sec and then regridded to our final grid spacing of 7.03125 km in the horizontal $x$ and $y$ directions and 5 km in the vertical $z$ direction. The simulation ran for another 1000~sec at this resolution with 5 second cadence.

To capture the center to limb variation effects more accurately,
the photospheric abundances of \cite{Asplund2009} and 12 band opacity were used in the simulation.

\subsection{Radiative transfer}
One of the developed snapshots of the simulation was used to generate the synthetic
G-band spectrum for a wavelength interval between 429 to 432~nm, using 3000 spectral points
with the 1D version of the radiative transfer code RH \citep{Uitenbroek2001} in local thermodynamic equilibrium.
Intensities were also synthesized at a single continuum wavelength of 500~nm.
To save computational time the vertical grid space was degraded to the 10~km by using only 
every second grid point along the vertical axes. We synthesized spectra at the disk and 
toward the limb at $\mu$\,=\,1 and 0.85, respectively, where $\mu$ is the cosine of heliocentric angle, 
$\theta$ (middle row of Figure~\ref{fig3}).

To synthesize G-band intensities at $\mu$ = 0.85, the MURaM cube was modified by shifting 
each horizontal layer relative to the layer below by $\Delta{z}\tan{\theta}$, where $\Delta{z}$ 
represents the vertical grid spacing. This adjustment aligns the slanted LoS direction vertically. 
Pixel sampling was foreshortened along the y-direction by a factor of $\sqrt{1-\mu^2}$ and increased along the $z$-direction 
by a factor of $1/\mu$ for the $\mu$ = 0.85 data.

To closely replicate the DKIST G-band filtergrams, the
synthetic data were multiplied by the VBI transmission
profile (bottom panel of Figure~\ref{fig3}) and integrated over the
whole wavelength range covered by the VBI filter. 

To investigate the polarization signatures of the intensity striation, we also synthesized the full 
Stokes profiles of the magnetically sensitive Fe {\sc{i}} 630.2~nm photospheric line, using 
the same MURaM cube at $\mu$ = 0.85 over the FoV shown in Figure~\ref{fig5}.


\section{Analysis and results}
\label{an_res}
\subsection{DKIST filtegrams}
\label{an_obs}

Figure~\ref{fig1} displays the G-band image of the plage region taken on 2022 June 3 at 18:36 UT. 
The plage is dominated by facular brightenings 
that are associated with strong magnetic flux concentrations (MFCs) \citep[see Figure~1 \& 2 in][]{Kuridze2024}.
The zoomed G-band image in the left panel of Fig.~\ref{fig2} shows all facular features, such as bright granular edges that are 
bordered by dark lanes on the disk-center side. The bright granular edges are dominated 
by a pattern of striations\,---\,fine-scale, short, dark features that are mostly parallel to each
other and extended towards the limb (Figure~\ref{fig2}).

To enhance the fine-scale details and structure of the striation, we applied a Gaussian-filter subtraction method to the images. 
In particular, images are blurred with two different Gaussian kernels (with $\sigma$\,=\,1 and 3) and subtracted from each other. 
The resulting image is displayed in the right panel of Figure~\ref{fig2}.  It shows that all facular brightening are 
striated with very thin and short features.
 
To quantify the width of the striae, we measured the full width at half maximum (FWHM) of the Gaussian 
fits to the intensity profiles along cross-cuts of individual striae. 
 Figure~\ref{fig4} shows the intensity profiles of 
eight randomly selected dark striae, along with their Gaussian fits. The FWHM of these profiles ranges from 20 to 53 km.
The classical theoretical resolution limit $(1.22\lambda/D)$ for DKIST/G-band is $\sim$20 km.

\subsection{Synthetic data}

Figure~\ref{fig3}, middle row of panels, presents synthetic G-band images for heliocentric angles $\mu$\,=\,1 
(left panel, DC) and $\mu$\,=\,85 (right panel, off DC), whereby the spectrally resolved G-band spectrum was multiplied 
with the VBI transmission profile and integrated over the wavelength range covered by the filter. All common 
G-band features seen in the observations are reproduced in the simulated images including magnetic bright points 
at DC and striated bright faculae and the dark lanes at $\mu = 85$. The bottom panel shows in blue the spectrally 
resolved G-band spectrum averaged over the full FoV of the simulation together and overplotted with the FTS 
solar spectrum \citep{Neckel1984} of the same wavelength range (orange). They agree with each other to great detail. 
The very good agreement of synthetic and observational data in terms of facular details and G-band spectrum 
suggests that the simulated data can serve as a powerful diagnostics of G-band intensities.

\subsection{Dark lanes and facular brightenings}

From the MURaM snapshot, we also calculated the height of the surface where the optical depth 
reaches unity at different wavelength positions along the G-band. The top panels of Figure~\ref{fig5} present a 
small FoV of the full synthetic G-band image and the corresponding map of the height where the optical depth is 
unity in the G-band continuum at 430.7~nm (hereafter referred to as $\tau=1$) 
for the heliocentric angle $\mu$\,=\,0.85. The map reveals that the formation 
heights of the $\tau=1$ surface in the dark lanes and bordering granules on the disk-center side are very similar 
(top row of Figure~\ref{fig5}). 
However, the areas next to dark lanes on the limb side, where facular brightenings 
appear in the G-band images, show the $\tau$\,=\,1 surface to lie deeper compared to other parts of granules  and the dark lanes (top panels of Figure~\ref{fig5}).

The bottom panels of Figure~\ref{fig5} display the vertical cuts through the temperature and LoS 
magnetic field strength along the blue dotted lines marked in the top panels.
Isothermal contours at 
$T$\,=\,6000~and~7000~K are overplotted in white.
Blue lines follow the $\tau=1$ height. The base of the photosphere (the zero point of the geometrical height scale, $z=0$) 
is defined as the average height where optical depth at 500~nm is unity at DC.

The diagrams in the bottom panels of Figure~\ref{fig5} confirm that the continuum intensity from the dark lane forms in a very similar 
geometrical height range as that from within the neighbouring granule on the disk-center side. LoSs that pass through the 
dark lane are all located on the disk-center side of the sharp edge where the $\tau=1$ surface drops into the MFC, however, 
they sense a distinctly lower temperature than in the surrounding areas further away, which makes them dark in G-band intensity. 
The low temperature comes about from enhanced radiative losses of the close surroundings into the MFC and the 
reduced temperature within the MFC compared to the surroundings at the same geometrical height.
At the edge of the granule next to the dark lane on the limb side, the LoS intersects a larger portion of the flux tube 
where density and with it opacity is reduced compared to the unperturbed atmosphere 
(bottom left panel of Figure~\ref{fig5}). Consequently, the $\tau=1$ surface forms deeper at the granular walls, 
where the temperature is higher ($\sim$7000 K) compared to the temperature within granular cells ($\sim$6000 K), 
resulting in the appearance of bright faculae (top left panel of Figure~\ref{fig5}).
The results are in agreement with the previous conclusions drawn from similar analyses  \citep{Keller2004,Carlsson2004,Steiner2005}.

\subsection{Striation}

Similar to the observations, striation along faculae is also apparent in the synthetic G-band images from the simulation at $\mu=0.85$ (top left panel of Figure~\ref{fig5}).
To quantify their widths in the simulated images, we examine the FWHM of the Gaussian fits to intensity profiles along cross-cuts 
of individual striae. Figure~\ref{fig6} displays the intensity profiles of eight randomly selected dark striae, with the 
corresponding Gaussian fits. The FWHM of these profiles ranges from 17 to 46 km, 
very similar to the FWHM from the observations given in Sect.~\ref{an_obs}.

We note that the intensity cuts used for the FWHM calculations are taken mostly perpendicular to both 
the observed and synthetic striae (Figure~\ref{fig4} \& \ref{fig6}). 
However, in some instances, due to local intensity variations, the cuts are 
slightly tilted from the perpendicular direction to avoid such local intensity variations for the Gaussian fit. 
Since the widths of the structures are very narrow (a few pixels), the slight tilt of the cross-cuts does not affect the FWHM estimates.

The top panels of Figure~\ref{fig5} show that the dark striae coincide with elevated $\tau=1$ heights. 
The difference in heights of the $\tau=1$ surface between the narrow dark features and their surrounding areas 
ranges for the instances 1--8 of Fig.~\ref{fig6} from 10 to 30 km.

Maps of Stokes parameters of the spectral line \ion{Fe}{1} 630.2~nm, synthesized with the same MURaM datacube at $\mu=0.85$, also
reveal striations that are co-spatial with those seen in the synthetic G-band images.
The magnetic field at the formation height of \ion{Fe}{1} 630.2~nm
is predominantly vertical with respect to the surface, 
making the vertical component ($B_{z}$ hereafter) the dominant one, while the $x$ and $y$ components can be neglected. 
For large heliocentric angles, Stokes $V$ is determined by $B_{z}$  and the viewing angle 
(i.e., the projection of $B_{z}$  along the LoS). At $\mu=0.85$, $B_{z}$ 
also has a strong transversal component with respect to the observer.
As a result, striations can also be seen in maps of Stokes $Q$ or $U$ (not shown here). 

Figure~\ref{fig7} displays the G-band intensity,  
$B_{z}$ (relative to the surface) at $z=0$, and the Stokes-$V$ signal
along the dashes crossing the striations. It demonstrates that the G-band striation is closely correlated 
with variations in $B_z$ and the Stokes-$V$ signal in the sense that $B_z$ is weaker in the dark striae than in the 
neighbouring bright ones.   Variation of $B_z$ associated 
with the striation ranges from approximately 100 to 250~G (Figure~\ref{fig7}).


\section{Discussion and conclusion}
\label{disc}

High-resolution G-band filtergrams obtained with DKIST, the largest solar optical telescope, reveal that facular 
regions 
exhibit a very thin and delicate striation pattern, consisting of dark and bright striae.
The width of these narrow features ranges from 
approximately 20 to 50 km (Figure~\ref{fig4}), indicating that they are below the resolution limit 
of 1-meter class solar telescopes. Moreover, 
some of the detections are close to the spatial resolution of the analyzed DKIST data ($\sim$0.03\arcsec), 
suggesting that the phenomenon most likely exist on even smaller spatial scales.

We successfully reproduced striated facular brightenings using a MURaM snapshot and radiative transfer 
calculations with spatial resolution and heliocentric angle comparable to the DKIST observations (Figure~\ref{fig3}).
The striation in the synthetic G-band images exhibit morphological characteristics, such as orientation, 
width, and length, similar to that of the observations (Figure~\ref{fig6}).
Similar to the observations, the width of some individual striae is around 17–20 km, which is close to 
the smallest possible scale in a simulation with 7~km grid spacing, meaning that these features could be even finer.

Analysis of numerical simulations and synthetic images confirms that the striations seen in 
filtergrams are linked to spatial variations 
in photospheric magnetic flux density, which cause shifts in the geometrical height where the emergent continuum 
intensity forms\,---\,a phenomenon known as the Wilson depression (Figure~\ref{fig5} \& \ref{fig7}). Some fine-scale striations 
in the synthetic images arise from variations in the magnetic field as small as 100 G and showing a 
Wilson depression as small as 10~km (top panels of Figure~\ref{fig5}). 

The origin of the magnetic field variations remain an open question, which requires further investigation. 
A proposed mechanism for creating such variation is the flute instability (also called "interchange instability"), 
which could lead to a corrugated magnetic surface on flux tubes \citep{Bunte1993,Sprut2010,Schlichenmaier2016}. This instability can develop
from perturbations at the flux tube boundary and can alter the tube’s shape (surface structure). Theoretical studies suggest that 
magnetic flux tubes in the photosphere near $\tau_{500\,\rm nm}=1$ can be unstable to the flute instability, leading to 
significant surface deformations \citep{Bunte1993,Sprut2010}.

Striations are also evident in \ion{Fe}{1} 630.2~nm Stokes-$V$ maps synthesized using the same 
MURaM datacube at $\mu=0.85$. Moreover, there is a close correlation between Stokes-$V$ and G-band signals along the dashes crossing the striations (Figure~\ref{fig7}). 
They have similar width and length in the synthetic Stokes-$V$ maps as their G-band counterparts. 
Therefore, high spatial resolution is also necessary to resolve them in magnetograms and/or polarization maps.
The excellent agreement between simulated and observed data suggests that DKIST G-band images trace variations of magnetic field strength 
and the footprints of Wilson depressions at similar scales. We note that the same plage 
region analyzed in this work has also been observed with the slit-based ViSP spectropolarimeter at DKIST, 
in the spectral channel that covers the magnetically sensitive photospheric lines \ion{Fe}{1} 630.1/630.2~nm 
\citep[see details in][]{Kuridze2024}. However, the resolution of the data ($\gtrsim0\farcs2$) was insufficient 
to resolve the striation in the magnetograms. Future DKIST spectropolarimetric observations with improved spatial 
resolution with ViSP, along with the upcoming Visible Tunable Filter (VTF) can target the magnetic striation.

Our analysis has also confirmed that facular brightenings are hot granular areas seen as 
a result of opacity reduction due to magnetic flux concentration of reduced mass density 
in front of them. The dark lanes form near the disk-center side edges of vertical magnetic 
flux concentrations viewed at an angle (Figure~\ref{fig5}).

We would like to add that striations have also been detected in DKIST/VBI observations in the 
blue-continuum channel at 450~nm at a large heliocentric angle ($\mu = 0.85$) near active region NOAA 1282 (not shown). 
We synthesized blue continuum images using MURaM snapshots and the RH code (not shown) at different heliocentric 
angles and found that the striation in this channel has very similar appearance and properties to that observed in the G-band.

The presented work demonstrates that facular striation provides a unique opportunity 
to observe spatial variations in magnetic flux density of very small spatial scale and amplitude with unprecedented detail. 
The small spatial scales and high sensitivity to magnetic flux variations suggest that striation may 
evolve on extremely short dynamical timescales. 
Investigating their temporal evolution and dynamical properties is beyond the scope of 
this paper. However, we plan to explore this in a follow-up study, including flute instability as 
a potential formation mechanism, using high spatial and temporal resolution data and simulations.
Additionally, we plan to synthesize G-band spectra and images from MURaM 
simulations with a grid spacing of 2$-$4 km to investigate whether 
striation appears at scales smaller than DKIST's spatial resolution limit.

We note that apart from faculae, striation is observed in other photospheric structures such as 
pores and sunspot light bridges \citep{Lites2004}. The fibrillar structure of the sunspot penumbra may 
have a similar origin as well. Therefore, similar high-resolution observational and modeling studies 
could be conducted to examine whether the findings presented in this work apply other type of 
photospheric striations as well. Furthermore, magnetically induced striation is also observed in more 
remote astrophysical objects such as molecular clouds, i.e., the Polaris flare molecular 
cloud \citep{skalidis+al2023} or the Taurus molecular cloud \citep{heyer+al2016}. 
Thus, we believe that the phenomenon is of general astrophysical interest 
and could even become the subject of laboratory experiments.

\begin{acknowledgments}
The research reported herein is based on data collected
with DKIST a facility of the National Science Foundation.
DKIST is operated by the National Solar Observatory under a
cooperative agreement with the Association of Universities for
Research in Astronomy, Inc. DKIST is located on land of
spiritual and cultural significance to Native Hawaiian people.
The use of this important site to further scientific knowledge is done so 
with appreciation and respect.
D.K. acknowledges the Georgian Shota Rustaveli National Science
Foundation project FR-22-7506. D.K. acknowledges the Science
and Technology Facilities Council (STFC) grant ST/W000865/.
This material is based upon work supported by the NSF National Center 
for Atmospheric Research, which is a major facility sponsored by the U.S. 
National Science Foundation under Cooperative Agreement No.\,1852977.
We would like to acknowledge high-performance computing support from 
the Derecho system (doi:10.5065/qx9a-pg09) provided by the NSF National 
Center for Atmospheric Research (NCAR), sponsored by the National Science Foundation.
\end{acknowledgments}
\bibliography{kuridze_et_al_2025}{}
\bibliographystyle{aasjournal}

\end{document}